\documentclass{article}
\usepackage{amsfonts}
\usepackage{amsmath}
\usepackage{amssymb}
\usepackage{amsthm}
\usepackage{fullpage}
\usepackage{graphicx}
\usepackage{ifthen}
\usepackage{color}

\definecolor{webgreen}{rgb}{0,.5,0}
\definecolor{webblue}{rgb}{0,0,.5}

\newtheorem{theorem}{Theorem}
\newtheorem{lemma}[theorem]{Lemma}
\newtheorem{conjecture}{Conjecture}

\DeclareMathOperator{\tr}{\textbf{tr}}

\newcommand{\sievefinal}{0}

\newcommand{\prelim}[1]{\ifthenelse{\equal{\sievefinal}{0}}{#1}{}}
\newcommand{\final}[1]{\ifthenelse{\equal{\sievefinal}{1}}{#1}{}}

\DeclareMathAlphabet{\varmathbb}{U}{bbold}{m}{n}
\newcommand{\one}{{\varmathbb 1}}
\renewcommand{\vec}[1]{\mathbf{#1}}

\newcommand{\remove}[1]{}

\newcommand{\C}{\mathbb{C}}

\newcommand{\Z}{\mathbb{Z}}

\newcommand{\U}{\textsf{U}}

\newcommand{\rk}{\textbf{rk}}
\newcommand{\reg}{\textbf{reg}}

\newcommand{\ket}[1]{\left| #1 \right\rangle}
\newcommand{\bra}[1]{\left\langle #1 \right|}
\newcommand{\inner}[2]{\left\langle #1, #2 \right\rangle}
\newcommand{\innerg}[2]{\left\langle #1, #2 \right\rangle_G}

\newcommand{\CG}{\C[G]}
\newcommand{\wg}{\widehat{G}}
\newcommand{\norm}[1]{\left\| #1 \right\|}
\newcommand{\abs}[1]{\left| #1 \right|}
\newcommand{\eps}{\epsilon}

\newcommand{\vsigma}{{\boldsymbol{\sigma}}}

\newcommand{\Root}{{\rm root}}
\newcommand{\planch}{{\mathcal P}_{\rm planch}}

\newcommand{\sub}{\scriptsize\mbox}
\newcommand{\prtriv}{P_T^{\{1\}}}
\newcommand{\prh}{P_T^H}
\newcommand{\rhotriv}{\rho_{\{1\}}}
\newcommand{\Plmt}{{P^{\lambda \otimes \mu}_\tau}}
\newcommand{\Pcoll}{P_{\rm coll}}

\newcommand{\Ind}[2]{{{\rm Ind}_{#1}^{#2}}}

\begin{document}

\title{On the Impossibility of a Quantum Sieve Algorithm\\ for Graph Isomorphism}

\author{Cristopher Moore\\ \texttt{moore@cs.unm.edu}\\ University of New Mexico 
\\ and the Santa Fe Institute \and Alexander Russell\\ \texttt{acr@cse.uconn.edu}\\ University of Connecticut  \final{\and Piotr \'{S}niady\\ \texttt{piotr.sniady@math.uni.wroc.pl}\\ University of Warsaw}}
\maketitle

\abstract{
It is known that any quantum algorithm for Graph Isomorphism that works within the framework of the hidden subgroup problem (HSP) must perform highly entangled measurements across $\Omega(n \log n)$ coset states.  One of the only known models for how such a measurement could be carried out efficiently is Kuperberg's algorithm for the HSP in the dihedral group, in which quantum states are adaptively combined and measured according to the decomposition of tensor products into irreducible representations.  This ``quantum sieve'' starts with coset states, and works its way down towards representations whose probabilities differ depending on, for example, whether the hidden subgroup is trivial or nontrivial.

In this paper we give strong evidence that no such approach can succeed for Graph Isomorphism.  Specifically, we consider the natural reduction of Graph Isomorphism to the HSP over the the wreath product $S_n \wr \Z_2$.  
\prelim{We show, modulo a group-theoretic conjecture regarding the asymptotic characters of the symmetric group, that no matter what rule we use to adaptively combine quantum states, there is a constant $b > 0$ such that no algorithm in this family can solve Graph Isomorphism in $e^{b \sqrt{n}}$ time.}
\final{We show that no matter what rule we use to adaptively combine quantum states, there is a constant $b > 0$ such that no algorithm in this family can solve Graph Isomorphism in $e^{b \sqrt{n}}$ time.} 
In particular, such algorithms are essentially no better than the best known classical algorithms, whose running time is $e^{O(\sqrt{n \log n})}$.
}

\section{Introduction}
\label{sec:intro}

Peter Shor's efficient quantum algorithms for order finding, factoring, and the discrete logarithm~\cite{Shor:1994:AQC}, and Simon's algorithm for determining the symmetry of a type of 2-1 function defined on $\{0,1\}^n$~\cite{Simon:1994:PQC}, led a frenzied charge to uncover the full algorithmic potential of a general purpose quantum computer. Creative invocations of the order-finding primitive yielded efficient quantum algorithms for a number of other number-theoretic problems~\cite{Hallgren:2002:PTQ,Hallgren:2005:FQA}. As the field matured, these algorithms were roughly unified under the general framework of the \emph{hidden subgroup problem}, where one must determine a subgroup $H$ of a group $G$ by querying an oracle $f: G \rightarrow S$ known to have the property that $f(g) = f(gh) \Leftrightarrow h \in H$. Solutions to this general problem are the foundation for almost all known superpolynomial speedups offered by quantum algorithms over their classical counterparts (see~\cite{Aharonov:2006:PQA} for an important recent exception).

The algorithms of Simon and Shor essentially solve the hidden subgroup problem on abelian groups, namely $\Z_2^n$ and $\Z_n^*$ respectively.  Since then, \emph{non-abelian} hidden subgroup problems have received a great deal of attention (e.g. \cite{Hallgren:2000:NSR,Grigni:2001:QMA,Friedl:2003:HTO,SODA::MooreRS2004,Bacon:2005:OMEA,Hallgren:2006:LQC}).  The cases where the ambient group $G$ is the dihedral group or the symmetric group possess the most compelling algorithmic applications: in particular, we can reduce Graph Isomorphism for rigid graphs of size $n$ to the case of the hidden subgroup problem over the symmetric group $S_{2n}$, or more specifically the wreath product $S_n \wr \Z_2$, where the hidden subgroup is promised to be either zero or of order two. 

The basic approach to these problems, motivated by the community's success with abelian groups, is to use the oracle to prepare ``coset states'' $\rho_H$ of the form
\[ \rho_H 
= \frac{1}{|G|} \sum_c \ket{cH}\bra{cH} \enspace,
\]
where $\ket{S}$, for a subset $S \subset G$, denotes the uniform superposition $(1/\sqrt{|S|}) \sum_g \ket{g}$. In the abelian case, one proceeds by computing the quantum Fourier transform of such coset states, measuring the resulting states, and appropriately interpreting the results. In the case of the symmetric group, however, determining $H$ from a quantum measurement of coset states is far more difficult.  In particular, no \emph{product measurement} (that is, a measurement which treats each coset state independently) can efficiently determine a hidden subgroup over $S_n$~\cite{Moore:2005:SGF}; in fact, any successful measurement must be \emph{entangled} over $\Omega(n \log n)$ coset states at once~\cite{Hallgren:2006:LQC}.

Unfortunately, our species' evolutionary history has given us little experience in thinking about highly-entangled quantum measurements.  So far, there are only two known families of such measurements which we can actually perform efficiently on a quantum computer:
\begin{description}
\item[\textbf{Implementation of the Pretty Good Measurement.}]  Pioneered by Bacon, Childs, and van Dam, this approach seeks to efficiently implement an explicit measurement, the \emph{pretty good measurement}.  For some choices of groups and subgroups, this measurement is known to be optimal~\cite{BaconCvD,Bacon:2005:OMEA,MooreR:PGM}; for arbitrary groups, it was recently shown to be sufficient when performed on $\Omega(\log |G|)$ coset states~\cite{HayashiKK06}.  On a number of groups, such as the affine, Heisenberg, and dihedral groups, implementing the pretty good measurement corresponds to solving random cases of a combinatorial problem.  However, it is not clear how to carry out this program for the symmetric groups, or even whether a corresponding combinatorial problem exists.
\item[\textbf{Sieves generated by Clebsch-Gordan decomposition.}]
Kuperberg's algorithm for the hidden subgroup problem in the dihedral group~\cite{Kuperberg} starts by generating a large number of coset states and subjecting each one to \emph{weak Fourier sampling}, so that it lies inside a known irreducible representation.  It then proceeds with an adaptive ``sieve'' process, at each step of which it judiciously selects pairs of states and measures them in a basis consistent with the \emph{Clebsch-Gordan} decomposition of their tensor product into irreducible representations.  This sieve continues until we obtain a state lying in an ``informative'' representation: namely, one from which information about the hidden subgroup can be easily extracted.  We can visualize the sieve as a forest, where leaves consist of coset states, each internal node measures the tensor product of its parents, and the informative representations lie at the roots.
\end{description}
The sieve approach is especially attractive in cases like Graph Isomorphism, where all we need to know is whether the hidden subgroup is trivial or nontrivial.  Specifically, suppose that the hidden subgroup $H$ is promised to be either the trivial subgroup $\{1\}$ or a \emph{conjugate} of a known subgroup $H_0$. Assume further that there is an irreducible representation $\sigma$ of $G$ with the property that $\sum_{h \in H_0} \sigma(h) = 0$; that is, a ``missing harmonic'' in the sense of~\cite{MooreR:banff}.  In this case, if $H$ is nontrivial then the probability of observing $\sigma$ under weak Fourier sampling of the coset state $\rho_H$ is zero.  More generally, the probability of observing $\sigma$ when measuring any state which is \emph{$H$-invariant}---that is, one which is fixed under right multiplication by any element of $H$---is zero.  As we discuss below, the sieve process maintains the $H$-invariance of the observed states throughout its operation, so it follows that $\sigma$ cannot appear at any time in the algorithm.  If, on the other hand, one can guarantee that this sieve process \emph{does} observe $\sigma$ with significant probability when the hidden subgroup is trivial and the corresponding states are completely mixed, it gives us an algorithm to distinguish the two cases.  

For example, if we consider the case of the hidden subgroup problem in the dihedral group $D_n$ where $H$ is either trivial or a conjugate of $H_0=\{1,m\}$ where $m$ is an involution, then the sign representation $\pi$ is a missing harmonic.  Applying Kuperberg's sieve, we observe $\pi$ with significant probability after $e^{O(\sqrt{n})}$ steps if $H$ is trivial, while we can never observe it if $H$ is of order $2$.  This framework has been successfully applied to groups of the form $G^n$ by Alagi\'{c} et al.~\cite{AlagicMR06}, suggesting that it might apply, mutatis mutandis, to a wide class of groups.

One difficulty in analyzing these sieve algorithms is that the algorithm designer has considerable flexibility in how she selects which pairs of representations to combine.  We will show below, however, that modulo a representation-theoretic conjecture on the symmetric group, the hidden subgroup problem related to Graph Isomorphism cannot be solved efficiently by any algorithm in this family. Specifically, no adaptive selection rule produces an efficient sieve whose transcript of observations statistically distinguishes the isomorphic and nonisomorphic cases, unless it combines $e^{\Omega(\sqrt{n})}$ coset states.  Since the best known classical algorithms for Graph Isomorphism run in time $e^{O(\sqrt{n \log n})}$~\cite{Babai:1980:CCL,Babai:1983:CLG}, quantum algorithms of this kind offer essentially no improvement over their classical counterparts.

The paper is organized as follows.  We give a brief review of group representation theory in Section~\ref{sec:representation-theory}.  In Section~\ref{sec:sieves} we precisely describe the family of sieve algorithms outlined above.  In Sections~\ref{sec:distributions} and~\ref{sec:trivial-and-legal}, we describe the probability distributions on the sequence of representations observed throughout a sieve algorithm.   In Sections~\ref{sec:wreath} and~\ref{sec:homogeneous}, we focus on wreath product groups of the form $G \wr \Z_2$, and show that the observed distributions completely fail to distinguish trivial and nontrivial subgroups unless we observe a special type of representation.   In Section~\ref{sec:collisions} we bound the probability of this event in terms of the characters of $G$.  Then, in Section~\ref{sec:conjecture}, we describe a conjectured bound on the characters of the symmetric group, and in Section~\ref{sec:main} we show that this conjecture implies our main result.  We conclude in Section~\ref{sec:evidence} by proving that our conjecture holds in a number of cases.

\section{Fourier analysis on finite groups}
\label{sec:representation-theory}

In this section we review the representation theory of finite groups. Our treatment is primarily for the purposes of setting down notation; we refer the reader to~\cite{Serre77} for a complete account.

Let $G$ be a finite group. A \emph{representation} $\sigma$ of $G$ is a homomorphism $\sigma: G \to \U(V)$, where $V$ is a finite-dimensional Hilbert space and $\U(V)$ is the group of unitary operators on $V$.  The \emph{dimension} of $\sigma$, denoted $d_\sigma$, is the dimension of the vector space $V$. 
Fixing a representation $\sigma: G \to \U(V)$, we say that a subspace $W \subset V$ is \emph{invariant} if $\sigma(g) \cdot W = W$ for all $g \in G$. 
When $\sigma$ has no invariant subspaces other than the trivial subspace $\{ \vec{0} \}$ and $V$ itself, $\sigma$ is said to be \emph{irreducible}. 

If two representations $\sigma$ and $\sigma'$ are the same up to a unitary change of basis, we say that they are \emph{equivalent}. It is a fact that any finite group $G$ has a finite number of distinct irreducible
representations up to equivalence and, for a group $G$, we let $\wg$ denote a set of representations containing exactly one from each equivalence class.  We often say that each $\sigma \in \wg$ is the \emph{name} of an irreducible representation, or an \emph{irrep} for short.

The irreps of $G$ give rise to the Fourier transform. Specifically, for a function $f: G \to \C$ and an element $\sigma \in \wg$, define the \emph{Fourier transform of $f$ at $\sigma$} to be
\[
\hat{f}(\sigma) = \sqrt{\frac{d_\sigma}{|G|}} \sum_{g \in G} f(g)\sigma(g)\enspace.
\]
The leading coefficients are chosen to the make the transform unitary, so that it preserves inner products:
\[
\langle f_1, f_2\rangle = \sum_g f_1^*(g)f_2(g)
= \sum_{\sigma \in \wg}\tr \!\left(\hat{f_1}(\sigma)^\dagger \cdot \hat{f_2}(\sigma)\right)\enspace.
\]

\remove{
then, there is a nontrivial invariant subspace $W \subset V$ and, as the inner product $\langle \cdot, \cdot\rangle$ is invariant under the unitary maps $\sigma(g)$, it is immediate that the dual subspace
$$
W^\perp = \{ \vec{u} \mid \forall \vec{w} \in W, \langle\vec{u},\vec{w}\rangle = 0\}
$$
is also invariant.  Associated with the decomposition $V = W \oplus W^\perp$ is the natural decomposition of the operators $\sigma(g) = \sigma_W(g) \oplus \sigma_{W^\perp}(g)$. By repeating this process, any representation $\sigma: G \to \U(V)$ may
} 

If $\sigma$ is \emph{not} irreducible, it can be decomposed into a direct sum of irreps $\tau_i$, each of which acts on an invariant subspace, and we write $\sigma \cong \tau_1 \oplus \cdots \oplus \tau_k$. 
In general, a given $\tau$ can appear multiple times in this decomposition, in the sense that $\sigma$ may have an invariant subspace isomorphic to the direct sum of $a_\tau$ copies of $\tau$.  In this case $a_\tau$ is called the \emph{multiplicity} of $\tau$ in the decomposition of $\sigma$.

There is a natural product operation on representations: if $\lambda: G \to \U(V)$ and $\mu: G \to \U(W)$ are representations of $G$, we may define a new representation $\lambda \otimes \mu: G \to \U(V \otimes W)$ as
$(\lambda \otimes \mu)(g): \vec{u} \otimes \vec{v} \mapsto \lambda(g)\vec{u} \otimes \mu(g)\vec{v}$.  This representation corresponds to the \emph{diagonal action} of $G$ on $V \otimes W$, in which we apply the same group element to both parts of the tensor product.  In general, the representation $\lambda \otimes \mu$ is not irreducible, even when both $\lambda$ and $\mu$ are. This leads to the \emph{Clebsch-Gordan
  problem}, that of decomposing $\lambda \otimes \mu$ into irreps.

Given a representation $\sigma$ we define the \emph{character} of $\sigma$,
denoted $\chi_\sigma$, to be the trace 
$\chi_\sigma(g) = \tr \sigma(g)$. As the trace of a linear operator is invariant
under conjugation, characters are constant on the conjugacy classes of $G$. 
Characters are a powerful tool for reasoning about
the decomposition of reducible representations. In particular, when
$\sigma = \bigoplus_i \tau_i$ we have $\chi_\sigma = \sum_i \chi_{\tau_i}$ and, moreover, for
$\sigma, \tau \in \wg$, we have the orthogonality conditions
$$
\innerg{\chi_\sigma}{\chi_\tau} = \frac{1}{|G|} \sum_{g \in G} \chi_\sigma(g)\chi_\tau(g)^*
= \begin{cases} 1 & \sigma = \tau\enspace,\\
  0 & \sigma \neq \tau\enspace. \end{cases}
$$
Therefore, given a representation $\sigma$ and an irrep $\tau$, the multiplicity $a_\tau$ with which $\tau $ appears in the decomposition of $\sigma$ is $\innerg{\chi_\tau}{\chi_\sigma}$.  For example, since $\chi_{\lambda \otimes \mu}(g) = \chi_\lambda(g) \cdot \chi_\mu(g)$,
the multiplicity of $\tau$ in the Clebsch-Gordan decomposition of $\lambda \otimes \mu$ is $\innerg{\chi_\tau}{\chi_\lambda \chi_\mu}$.

A representation $\sigma$ is said to be \emph{isotypic} if the irreducible factors appearing in the decomposition are all isomorphic, which is to say that there is a single nonzero $a_\tau$ in the decomposition above. Any representation $\sigma$ may be uniquely decomposed into maximal isotypic subspaces, one for each irrep $\tau$ of $G$; these subspaces are precisely those spanned by all copies of $\tau$ in $\sigma$. In fact, for each $\tau$ this subspace is the image of an explicit projection operator $\Pi_\tau$ which can be written 
\[ \Pi_\tau = \frac{1}{|G|} \sum_{g \in G} d_\tau \chi_\tau(g)^* \sigma(g) \enspace .  \]
A useful fact is that $\Pi_\tau$ commutes with the group action; that is, for any $h \in G$ we have
\[ \sigma(h) \Pi_\tau \sigma(h)^\dagger 
 = \frac{1}{|G|} \sum_{g \in G} d_\tau \chi_\tau(g)^* \sigma(hgh^{-1}) 
 = \frac{1}{|G|} \sum_{g \in G} d_\tau \chi_\tau(h^{-1} g h)^* \sigma(g) 
 = \frac{1}{|G|} \sum_{g \in G} d_\tau \chi_\tau(g)^* \sigma(g) 
= \Pi_\tau \enspace .
\]

Our algorithms will perform measurements which project into these maximal isotypic subspaces and observe the resulting irrep name $\tau$.  For the particular case of coset states, this measurement is called \emph{weak Fourier sampling} in the literature; however, since we are interested in a more general process which in fact performs a kind of strong multiregister sampling on the original coset states, we will use the term \emph{isotypic sampling} instead.

Finally, we discuss the structure of a specific representation, the \emph{(right) regular} representation $\reg$, which plays an important role in the analysis below. $\reg$ is given by the permutation action of $G$ on itself.  Specifically, let $\CG$ be the \emph{group algebra} of $G$; this is the $|G|$-dimensional vector space of formal sums
$$
\Bigl\{ \sum_g \alpha_g \cdot g \mid \alpha_g \in \C \Bigr\} \enspace .
$$
(Note that $\CG$ is precisely the Hilbert space of a single register containing a superposition of group elements.) Then $\reg$ is the representation $\reg: G \to \U(\CG)$ given by linearly extending right multiplication, $\reg(g): h \mapsto hg$.  It is not hard to see that its character
$\chi_\reg$ is given by
$$
\chi_\reg(g) = \begin{cases} |G| & g = 1\enspace,\\
  0 & g \neq 1 \enspace ,
\end{cases}
$$
in which case we have $\innerg{\chi_\reg}{\chi_\sigma} = d_\sigma$ for each $\sigma \in \wg$.  Thus $\reg$ contains $d_\sigma$ copies of each irrep $\sigma \in \wg$, and counting dimensions on each side of this decomposition implies 
\begin{equation}
\label{eq:planch}
|G| = \sum_{\sigma \in \wg} d_\sigma^2 \enspace.
\end{equation}
This equation suggests a natural probability distribution on $\wg$, the \emph{Plancherel distribution}, which assigns to each irrep $\sigma$ the probability $\planch^G(\sigma) = d_\sigma^2 / |G|$.  This is simply the dimensionwise fraction of $\CG$ consisting of copies of $\sigma$; indeed, if we perform isotypic sampling on the completely mixed state on $\CG$, or equivalently the coset state where the hidden subgroup is trivial, we observe exactly this distribution.

In general, we can consider subspaces of $\CG$ that are invariant under
left multiplication, right multiplication, or both; these subspaces are called
\emph{left-}, \emph{right-}, or \emph{bi-invariant} respectively.  
For each $\sigma \in \wg$, the maximal $\sigma$-isotypic subspace is a $d_\sigma^2$-dimensional bi-invariant subspace; it can be broken up further into $d_\sigma$ $d_\sigma$-dimensional
left-invariant subspaces, or (transversely) $d_\sigma$ $d_\sigma$-dimensional
right-invariant subspaces.  However, this decomposition is not unique.
If $\sigma$ acts on a vector space $V$, then choosing an orthonormal
basis for $V$ allows us to view $\sigma(g)$ as a $d_\sigma \times d_\sigma$ matrix.
Then $\sigma$ acts on the $d_\sigma^2$-dimensional space
of such matrices by left or right multiplication, and the columns and rows correspond
to left- and right-invariant spaces respectively.

\section{Clebsch-Gordan sieves}
\label{sec:sieves}

Consider the hidden subgroup problem over a group $G$ with the added promise that the hidden subgroup $H$ is either the trivial subgroup, or a conjugate of some fixed nontrivial subgroup $H_0$.  We shall consider sieve algorithms for this problem that proceed as follows:
\begin{enumerate}
\item The oracle is used to generate $\ell=\ell(n)$ coset states $\rho_H$, each of which is subjected to weak Fourier sampling. 
This results in a set of states $\rho_i$, where $\rho_i$ is a mixed state known to lie in the maximal $\sigma_i$-isotypic subspace of $\C[G]$ for some irrep $\sigma_i$.  
\item The following \emph{combine-and-measure} procedure is then repeated as many times as we like. 
Two states $\rho_i$ and $\rho_j$ in the set are selected according to an arbitrary adaptive rule that may depend on the entire history of the computation (in existing algorithms of this type, this selection in fact depends only on the irreps $\sigma_i$ and $\sigma_j$ in which they lie).  We then perform isotypic sampling on their tensor product $\rho_i \otimes \rho_j$: that is, we apply a measurement operator which observes an irrep $\sigma$ in the Clebsch-Gordan decomposition of $\sigma_i \otimes \sigma_j$ (see~\cite{Kuperberg} or~\cite{MooreR:banff} for how this measurement can actually be carried out by applying the diagonal action).  This measurement destroys $\rho_i$ and $\rho_j$, and results in a new mixed state $\rho$ which lies in the maximal $\sigma$-isotypic subspace; we add this new state to the set.
\item Finally, depending on the sequence of observations resulting from the measurements above, the algorithm guesses the hidden subgroup.
\end{enumerate}

We set down some notation to discuss the result of applying such an algorithm. Fixing a group $G$ and a subgroup $H$, let $A$ be a sieve algorithm which initially generates $\ell$ coset states.  As a bookkeeping tool, we will describe intermediate states of $A$'s progress as a \emph{forest of labeled binary trees}.  Throughout, we will maintain the invariant that the roots of the trees in this forest correspond to the current set of states available to the algorithm.  

Initially, the state of the algorithm consists of a forest consisting of $\ell$ single-node trees, each of which is labeled with the irrep name $\sigma_i$ that resulted from performing weak Fourier sampling on a coset state, and is associated with the resulting state $\rho_i$.  Then, each combine-and-measure step selects two root nodes, $r_1$ and $r_2$, and applies isotypic sampling to the tensor product of their states.  We associate the resulting state $\rho$ with a new root node $r$, and place the nodes $r_1$ and $r_2$ below it as its children.  We label this new node\footnote{New nodes is good nodes.} with the irrep name $\sigma$ observed in this measurement.

Thus, every node of the forest corresponds to a state that existed at some point during the algorithm, and each node $i$ is labeled with the name of the irrep $\sigma_i$ observed in the isotypic measurement performed when that node was created.  We call the resulting labeled forest the \emph{transcript} of the algorithm: note that this transcript contains all the information the algorithm may use to determine the hidden subgroup.

We make several observations about algorithms of this type.  First, it is easy to see that nothing is gained by combining $t > 2$ states at a time; we can simulate this with an algorithm which builds a binary tree with $t$ leaves, and which ignores the results of all its measurements except the one at the root.

Second, the algorithm maintains the following kind of symmetry under the action of the subgroup $H$.  Suppose we have a representation $\sigma$ acting on a Hilbert space $V$.  Given a subgroup $H$, we say that a state $\psi \in V$ is \emph{$H$-invariant} if $\sigma(h) \cdot \psi = \psi$ for all $h \in H$.  Similarly, given a mixed state $\rho$, we say that $\rho$ is $H$-invariant if $\sigma(h) \cdot \rho \cdot \sigma(h)^\dagger = \rho$ or, equivalently, if $\sigma(h)$ and $\rho$ commute.  For instance, the coset state $\rho_H$ is $H$-invariant under the right regular representation, since right-multiplying by any $h \in H$ preserves each left coset $cH$.  Now, suppose that $\rho_1$ and $\rho_2$ are $H$-invariant; clearly $\rho_1 \otimes \rho_2$ is $H$-invariant under the diagonal action, and performing isotypic sampling preserves $H$-invariance since $\Pi_\tau$ commutes with the action of any group element.  Thus the states produced by the algorithm are $H$-invariant throughout.

Third, it is important to note that while at each stage we observe only an irrep name, rather than a basis vector inside that representation, by iterating this process the sieve algorithm actually performs a kind of \emph{strong multiregister Fourier sampling} on the original set of coset states.  For instance, in the dihedral group, suppose that performing weak Fourier sampling on two coset states results in the two-dimensional irreps $\sigma_j$ and $\sigma_k$, and that we then observe the irrep $\sigma_{j+k}$ under isotypic sampling of their tensor product.  We now know that the original coset states were in fact confined to a particular subspace, spanned by two entangled pairs of basis vectors.

Finally, we note that the states produced by a sieve algorithm are quite different from coset states.  In particular, they belong not to a maximal isotypic subspace of $\CG$, but to a (typically much higher-dimensional)  \emph{non-maximal} isotypic subspace of $\CG^{\otimes \ell}$, where $\ell$ is the number of coset states feeding into that state (i.e., the number of leaves of the corresponding tree).  Moreover, they have more symmetry than coset states, since each isotypic measurement implies a symmetry with respect to the diagonal action on the set of leaves descended from the corresponding internal node.  In the next sections we will show how these states can be written in terms of projection operators applied to this high-dimensional space.

\remove{
One special case in which sieve algorithms are attractive is when the nontrivial subgroup $H_0$ possesses a \emph{missing harmonic}, i.e., an irrep $\tau$ such that $\sum_{h \in H_0} \tau(h) = 0$.  Then as shown in~\cite{MooreR:banff}, we can never observe $\tau$ at any point in the sieve; and so if we do observe it, we know that the hidden subgroup is trivial.  This is how, for instance, the sieve algorithm for groups of the form $G^n$ in~\cite{AlagicMR06} works.  However, the negative results we present here do not rely on the notion of a missing harmonic.
}

\section{Observed distributions for fixed topologies}
\label{sec:distributions}

In general, the probability distributions arising from the combine-and-measure steps of a sieve algorithm depend on both the hidden subgroup and the entire history of previous measurements and observations (that is, the labeled forest, or transcript, describing the algorithm's history thus far). In this section and the next, we focus on the probability distribution induced by a \emph{fixed forest topology} and subgroup $H$.  We can think of this either as the probability distribution conditioned on the forest topology, or as the distribution of transcripts produced by some \emph{non-adaptive} sieve algorithm, which chooses which states it will combine and measure ahead of time.  We will show, modulo our conjecture, that for all forest topologies of sufficiently small size, the induced distributions on irrep labels fail to distinguish trivial and nontrivial subgroups.  Then, in Section~\ref{sec:main}, we will complete the argument for adaptive algorithms.

Clearly, in this non-adaptive case the distributions of irrep labels associated with different trees in the  forest are independent.  Therefore, we can focus on the distribution of labels for a specific tree. At the leaves, the labels are independent and identically distributed according to the distribution resulting from weak Fourier sampling a coset state~\cite{Hallgren:2000:NSR}, namely 
\[
\mathcal{D}_H(\sigma) = \frac{d_\sigma |H|}{|G|} \rk \left( \sum_{h \in H} \sigma(h)\right)\enspace.
\]
However, as we move inside the tree and condition on the irrep labels observed previously, the resulting distributions are quite different from this initial one.  To calculate the resulting joint probability distribution, we need to define projection operators acting on $\CG^{\otimes \ell}$ corresponding to the isotypic measurement at each node.

First, note that the coset state $\rho_H$ can be written in the following convenient form:
\[ \rho_H 
= \frac{1}{|G|} \sum_c \ket{cH}\bra{cH} 
= \frac{1}{|G|} \sum_{h \in H} \reg(h) \]
where $\reg$ is the \emph{right regular representation} defined in Section~\ref{sec:intro}: that is, $\rho_H$ is proportional to the projection operator which right-multiplies by a random element of $H$, 
\[ \Pi_H = \frac{1}{|H|} \sum_{h \in H} \reg(h) \enspace . \]
If $H$ is trivial, $\rho_H$ is the completely mixed state $\rhotriv = (1/|G|) \one$.  On the other hand, if $H=\{1,m\}$ for an involution $m$, we have 
\[ \rho_H 
= \frac{1}{|G|} (1 + \reg(m)) \]
which is proportional to the projection operator
\[ \Pi_H = \frac{1}{2} (1+\reg(m)) \enspace . \]

Now consider the tensor product of $\ell$ ``registers,'' each containing a coset state.  Given a linear operator $M$ on $\C[G]$ and a subset $I \subseteq [\ell] = \{1,\ldots,\ell\}$, let $M^I$ 
denote the operator on $\C[G^\ell] \cong \C[G]^{\otimes \ell}$ which applies $M$ to the registers in $I$ and leaves the other registers unchanged.  Then the mixed state consisting of $\ell$ independent coset states can be written in a number of ways, 
\[ 
\rho_H^{\otimes \ell} 
= \frac{1}{|G|^\ell} \prod_{j=1}^\ell (1+\reg(m)^{\{j\}}) 
= \frac{1}{|G|^\ell} \sum_{I \subseteq [\ell]} \reg(m)^I 
\]
and this state is proportional to a projection operator on $\C[G^\ell]$,
\begin{equation}
\label{eq:h} 
\Pi_H^{\otimes \ell}
= \frac{1}{2^\ell} \prod_{j=1}^\ell (1+\reg(m)^{\{j\}}) 
= \frac{1}{2^\ell} \sum_{I \subseteq [\ell]} \reg(m)^I 
\enspace. 
\end{equation}
Note the sum over subsets of registers, a theme which has appeared repeatedly in discussions of multiregister Fourier sampling~\cite{BaconCvD,Hallgren:2006:LQC,Kuperberg,MooreR:banff,MooreR:part2,Regev:2002:QCL}.

Now consider a tree $T$ with $\ell$ leaves corresponding to the $\ell$ initial registers, and $k$ nodes including the leaves.  We can represent this tree as a set system, in which each node $i$ is associated with the subset $I_i \subseteq [\ell]$ of leaves descended from it.  In particular, $I_\Root = [\ell]$ and $I_j=\{ j \}$ for each leaf $j$.  

Performing isotypic sampling at a node $i$ corresponds to applying the diagonal action to its children (or in terms of the algorithm, its parents) and inductively to the registers in $I_i$: that is, we multiply each register in $I_i$ by the same element $g$ and leave the others fixed.   If $\sigma_i$ is the irrep label observed at that node, let us denote its character and dimension by $\chi_i$ and $d_i$ respectively, rather than the more cumbersome $\chi_{\sigma_i}$ and $d_{\sigma_i}$.  Then the projection operator corresponding to this observation is
\begin{equation}
\label{eq:node}
 \Pi^T_i = \frac{1}{|G|} \sum_{g \in G} d_i \chi_i(g)^* \,\reg(g)^{I_i} 
\end{equation} 
Now consider a transcript of the sieve process which results in observing a set of irrep labels $\vsigma = \{\sigma_i\}$ on the internal nodes of the tree.  The projection operator associated with this outcome is 
\begin{equation}
\label{eq:pi_t}
\Pi^T[\vsigma] = \prod_{i=1}^k \Pi^T_i \enspace . 
\end{equation}
We will abbreviate this as $\Pi^T$ whenever the context is clear.  
Note that the various $\Pi^T_i$ in the product~\eqref{eq:pi_t} pairwise commute, since for any two nodes $i, j$ either $I_i$ and $I_{j}$ are disjoint, or one is contained in the other.  In the former case $a^{I_i}$ and $b^{I_{j}}$ for all $a,b$.  In the latter case, say if $I_i \subset I_{j}$, we have $a^{I_i} b^{I_{j}} = b^{I_{j}} (b^{-1} a b)^{I_i}$, and since $\chi_i(b^{-1} a b) = \chi_i(a)$ it follows from~\eqref{eq:node} that $\Pi^T_i \Pi^T_{j} = \Pi^T_{j} \Pi^T_i$.

Given a tree $T$ with $k$ nodes, we write $\prtriv[\vsigma]$ for the probability that we observe the set of irrep labels $\vsigma=\{\sigma_i\}$ in the case where the hidden subgroup is trivial.  Since the tensor product of coset states is then the completely mixed state in $\C[G^\ell]$, this is simply the dimensionwise fraction of $\C[G^\ell]$ consisting of the image of $\Pi^T$, or
\[
\prtriv[\vsigma] = \frac{1}{|G|^\ell} \tr \Pi^T \enspace . 
\]
Moreover, since measuring a completely mixed state results in the completely mixed state in the observed subspace, each state produced by the algorithm is completely mixed in the image of $\Pi^T$.  In particular, if the irrep label at the root of a tree is $\sigma$, the corresponding state consists of a classical mixture across some number of copies of $\sigma$, in each of which it is completely mixed.  Thus, when combining two parent states with irrep labels $\lambda$ and $\mu$, we observe each irrep $\tau$ with probability equal to the dimensionwise fraction of $\lambda \otimes \mu$ consisting of copies of $\tau$, namely
\begin{equation}
\label{eq:ptau}
 \Plmt = \frac{d_\tau}{d_\lambda d_\mu} \inner{\chi_\tau}{\chi_\lambda \chi_\mu}_G
\end{equation}
(recall that $\inner{\chi_\tau}{\chi_\lambda \chi_\mu}_G$ is the multiplicity of $\tau$ in $\lambda \otimes \mu$).    We will refer to this as the \emph{natural distribution} in $\lambda \otimes \mu$.

Now let us consider the case where the hidden subgroup is nontrivial.  Since the mixed state $\rho_{H^\ell}$ can be thought of as a pure state chosen randomly from the image of $\Pi_H^{\otimes \ell}$, the probability of observing a set of irrep labels $\vsigma$ in this case is
\[ \prh[\vsigma] = \frac{\tr \Pi^T \Pi_H^{\otimes \ell}}{\tr \Pi_H^{\otimes \ell}} = \frac{2^\ell}{|G|^\ell} \tr \Pi^T \Pi_H^{\otimes \ell} \]
where we use the fact that $\tr \Pi_H^{\otimes \ell} = [G:H]^\ell = (|G|/2)^\ell$.  Below we abbreviate these distributions as $\prtriv$ and $\prh$ whenever the context is clear.  Our goal is to show that, until the tree $T$ is deep enough, these two distributions are extremely close, so that 
the algorithm fails to distinguish subgroups of the form $\{1,m\}$ from the trivial subgroup.

\section{Trivial and legal assignments}
\label{sec:trivial-and-legal}

In this section we derive explicit expressions for $\prtriv$ and $\prh$.  First, we fix some additional notation.  Given an assignment of group elements $\{a_i\}$ to the nodes, for each leaf $j$ we let  $\prod_{i \leadsto j} a_i$ denote the product of the elements along the path from the root to $j$:
\[ \prod_{i \leadsto j} a_i = \prod_{i:j \in I_i} a_i \]
where the product is taken in order from to the root to the leaf.  Then using~\eqref{eq:node} and~\eqref{eq:pi_t} we can write
\begin{equation}
\label{eq:piT}
\Pi^T = \frac{1}{|G|^k} \sum_{\{a_i\}} d_i \chi_i(a_i)^* 
\bigotimes_{j=1}^\ell \reg\!\left( \prod_{i \leadsto j} a_i \right)^{\!\{i\}} \enspace . 
\end{equation}
We say that an assignment $\{a_i\}$ is \emph{trivial} if $\prod_{i \leadsto j} a_i = 1$ for every leaf $j$.  Then, since $\tr \reg(g) = \chi_\reg(g) = |G|$ if $g=1$ and $0$ otherwise, we have
\begin{equation}
\label{eq:prtriv}
\prtriv = \frac{1}{|G|^\ell} \tr \Pi^T
= \frac{1}{|G|^k} \sum_{\{a_i\} \sub{ trivial }} \prod_{i=1}^k d_i \chi_i(a_i)^* \enspace . 
\end{equation}
To get a sense of how this expression scales, note that the particular trivial assignment where $a_i = 1$ for all $i$ contributes $\prod_{i=1}^k d_i^2 / |G| = \prod_i \planch(\sigma_i)$, as if the $\sigma_i$ were independent and Plancherel-distributed.

Now let us consider $\prh$.  Combining~\eqref{eq:h} with~\eqref{eq:piT} gives the following expression for $\Pi^T \Pi_H^{\otimes \ell}$: 
\begin{equation}
\label{eq:prh}
\Pi^T \Pi_H^{\otimes \ell} = \frac{1}{2^\ell |G|^k} \sum_{\{a_i\}} d_i \chi_i(a_i)^* 
\bigotimes_{j=1}^\ell \reg\!\left( \left( \prod_{i \leadsto j} a_i \right) (1 + m) \right)^{\!\{i\}} \enspace . 
\end{equation}
We say that an assignment $\{a_i\}$ is  \emph{legal} if $\prod_{i \leadsto j} a_i \in \{1,m\}$ for every leaf $j$.  Then the trace of the term corresponding to $\{a_i\}$ is $|G|^\ell$ if $\{a_i\}$ is legal, and is $0$ otherwise, and analogous to~\eqref{eq:prtriv} we have
\begin{equation}
\label{eq:prh2}
\prh = \frac{2^\ell}{|G|^\ell} \tr \Pi^T \Pi_H^{\otimes \ell} 
= \frac{1}{|G|^k} \sum_{\{a_i\} \sub{ legal }} \prod_{i=1}^k d_i \chi_i(a_i)^* \enspace . 
\end{equation}
Thus these two distributions differ exactly by the terms corresponding to assignments which are legal but nontrivial.  Our main result will depend on the fact that for most $\vsigma$ these terms are identically zero, in which case $\prh$ and $\prtriv$ coincide.

\section{The wreath product and its representations}
\label{sec:wreath}

For any group $G$, the wreath product $G \wr \Z_2$ is the semidirect product $(G \times G) \rtimes \Z_2$, where we extend $G \times G$ by an involution which exchanges the two copies of $G$.  Thus the elements $((\alpha,\beta),0)$ form a normal subgroup $K \cong G \times G$ of index 2, and the elements $((\alpha,\beta),1)$ form its nontrivial coset.  We will call these elements ``non-flips'' and ``flips,'' respectively.  

The Graph Isomorphism problem reduces to the hidden subgroup problem on $S_n \wr \Z_2$ in the following natural way.  We consider the disjoint union of the two graphs, and consider permutations of their $2n$ vertices.  Then $S_n \wr \Z_2$ is the subgroup of $S_{2n}$ which either maps each graph onto itself (the non-flips) or exchanges the two graphs (the flips).  We assume for simplicity that the graphs are rigid.  Then if they are nonisomorphic, the hidden subgroup is trivial; if they are isomorphic, $H=\{1,m\}$ where $m$ is a flip of the form $((\alpha,\alpha^{-1}),1)$, where $\alpha$ is the permutation describing the isomorphism between them.  Note that the set of such flips constitutes a conjugacy class. 

For any group $G$, the irreps of $G \wr \Z_2$ can be written in a simple way in terms of the irreps of $G$.  It is useful to construct them by \emph{inducing} upward from the irreps of $K \cong G \times G$ (see~\cite{Serre77} for the definition of an induced representation).  First, each irrep of $K$ is the tensor product $\lambda \otimes \mu$ of two irreps of $G$.  Inducing this irrep from $K$ up to $G$ gives a representation
\[ \sigma_{\{\lambda,\mu\}} = \Ind{K}{G} (\lambda \otimes \mu) \]
of dimension $2 d_\lambda d_\mu$.  If $\lambda \not\cong \mu$, then this is irreducible, and $\sigma_{\{\lambda,\mu\}} \cong \sigma_{\{\mu,\lambda\}}$ (hence the notation).  We call these irreps \emph{inhomogeneous}.  Their characters are given by
\begin{equation}
\label{eq:inhomchar}
\chi_{\{\lambda,\mu\}}((\alpha,\beta),t) 
= \begin{cases} 
  \chi_\lambda(\alpha) \chi_\mu(\beta) + \chi_\mu(\alpha) \chi_\lambda(\beta) & \text{if}\;t=0 \\
  0 & \text{if}\;t=1 
\end{cases} \enspace .
\end{equation}
In particular, the character of an inhomogeneous irrep is zero at any flip.

On the other hand, if $\lambda \cong \mu$, then $\sigma_{\{\lambda,\lambda\}}$ decomposes into two irreps of dimension $d_\lambda^2$, which we denote $\sigma_{\{\lambda,\lambda\}}^+$ and $\sigma_{\{\lambda,\lambda\}}^-$.  We call these irreps \emph{homogeneous}.  Their characters are given by
\begin{equation}
\label{eq:homchar}
\chi_{\{\lambda,\lambda\}}^\pm((\alpha,\beta),t)
= \begin{cases} 
  \chi_\lambda(\alpha) \chi_\lambda(\beta) & \text{if}\;t = 0 \\
  \pm \chi_\lambda(\alpha\beta) & \text{if}\;t=1 
  \end{cases}  \enspace . 
\end{equation}
In the next section, we will show that sieve algorithms obtain precisely zero information that distinguishes hidden subgroups of the form $\{1,m\}$ from the trivial subgroup until it observes at least one homogeneous representation.

\section{The importance of being homogeneous}
\label{sec:homogeneous}

Suppose that the irrep labels $\vsigma = \{ \sigma_i \}$ observed during a run of the sieve algorithm consist entirely of inhomogeneous irreps of $G \wr \Z_2$.  Since the irreps have zero character at any flip, the only trivial or legal assignments $\{ a_i \}$ that contribute to the sums~\eqref{eq:prtriv} and~\eqref{eq:prh2} are those where each $a_i$ is a non-flip, i.e., is contained in the subgroup $K \cong G \times G$.  But the product of any string of such elements is also contained in $K$, so if this product is in $H=\{1,m\}$ where $m \notin K$, it is equal to $1$.  Thus any legal assignment of this kind is trivial, the sums~\eqref{eq:prtriv} and~\eqref{eq:prh2} coincide, and the probability of observing $\vsigma$ is the same in the trivial and nontrivial cases. That is, so long as every $\sigma_i$ in $\vsigma$ is inhomogeneous,
\begin{equation}
\label{eq:equality}
P^H_T[\vsigma] = P^{\{1\}}_T[\vsigma] \enspace .
\end{equation}
\remove{
This implies the following bound on the total variation distance between the two distributions, 
\begin{equation}
\label{eq:same}
 \frac{1}{2} \norm{\prh-\prtriv}_1 
 \le \Pr[ \text{at least one homogeneous irrep $\sigma$ has been observed in $T$} \mid H=\{1\} ] 
 \enspace . 
\end{equation}
}
Our strategy will be to show that observing even a single homogeneous irrep is unlikely, unless the tree generated by the sieve algorithm is quite large.  Moreover, because the two distributions coincide unless this occurs, it suffices to show that this is unlikely in the case where $H$ is trivial.

Now, it is easy to see that the probability of observing a given representation in $G \wr \Z_2$, under either the Plancherel distribution or a natural distribution, factorizes neatly into the probabilities that we observe the corresponding pair of irreps, in either order, in a pair of similar experiments in $G$.  First, the Plancherel measure of an inhomogeneous irrep $\sigma_{\{\lambda,\mu\}}$ is 
\begin{equation}
\label{eq:planch-wreath-inhom}
 \planch^{G \wr \Z_2}(\sigma_{\{\lambda,\mu\}})
= \frac{(2 d_\lambda d_\mu)^2}{2 |G|^2}
= 2 \,\planch^G(\lambda) \,\planch^G(\mu) \enspace . 
\end{equation}
Similarly, the probability that we observe a homogeneous irrep $\sigma^\pm_{\{\lambda,\lambda\}}$ is the probability of observing $\lambda$ twice under the Plancherel distribution in $G$, in which case the sign $\pm$ is chosen uniformly:
\begin{equation}
\label{eq:planch-wreath-hom}
 \planch^{G \wr \Z_2}(\sigma^\pm_{\{\lambda,\lambda\}})
= \frac{d_\lambda^4}{|G|^2} 
= \planch^G(\lambda)^2 \enspace .
\end{equation}

Now consider the natural distribution in the tensor product of two inhomogeneous irreps $\sigma_{\{\lambda,\lambda'\}}$ and $\sigma_{\{\mu,\mu'\}}$.  The multiplicity of a given homogeneous irrep $\sigma^\pm_{\{\tau,\tau\}}$ in this tensor product factorizes as follows, 
\[ \inner{\chi_{\{\tau,\tau\}}^\pm}{\chi_{\{\lambda,\lambda'\}} \chi_{\{\mu,\mu'\}}}_{G \wr \Z_2} 
= \frac{1}{2} 
\left( \innerg{\chi_\tau}{\chi_\lambda \chi_\mu} \innerg{\chi_\tau}{\chi_{\lambda'} \chi_{\mu'}} 
+ \innerg{\chi_\tau}{\chi_\lambda \chi_{\mu'}} \innerg{\chi_\tau}{\chi_{\lambda'} \chi_\mu} \right)
\enspace . 
\]
Thus the probability of observing either $\sigma_{\{\tau,\tau\}}^+$ or $\sigma_{\{\tau,\tau\}}^-$ under the natural distribution is
\begin{equation}
\label{eq:prob-hom}
P^{\sigma_{\{\lambda,\lambda'\}} \otimes \sigma_{\{\mu,\mu'\}}}_{\sigma_{\{\tau,\tau\}}^\pm}
= \frac{1}{2} \left( P^{\lambda \otimes \mu}_\tau P^{\lambda' \otimes \mu'}_\tau 
+ P^{\lambda \otimes \mu'}_\tau P^{\lambda' \otimes \mu}_\tau \right) \enspace . 
\end{equation}
In other words, the probability of observing a homogeneous irrep of $G \wr \Z_2$ is the probability of observing the same irrep in two natural distributions on $G$.  Let us denote the probability that we observe the same irrep in the natural distributions in $\lambda \otimes \mu$ and $\lambda' \otimes \mu'$---that is, that these two distributions collide---as
\[ \Pcoll^{\lambda \otimes \mu, \lambda' \otimes \mu'} =
 \sum_\tau P^{\lambda \otimes \mu}_\tau P^{\lambda' \otimes \mu'}_\tau \enspace . \]
 Then~\eqref{eq:prob-hom} implies that the total probability of observing a homogeneous irrep is 
 \begin{equation}
 \label{eq:prob-hom-bound}
\sum_\tau P^{\sigma_{\{\lambda,\lambda'\}} \otimes \sigma_{\{\mu,\mu'\}}}_{\sigma_{\{\tau,\tau\}}^\pm}
= \frac{1}{2} \left( \Pcoll^{\lambda \otimes \mu, \lambda' \otimes \mu'}
+ \Pcoll^{\lambda \otimes \mu', \lambda' \otimes \mu} \right) 
 \le \max \left( \Pcoll^{\lambda \otimes \mu, \lambda' \otimes \mu'},
  \Pcoll^{\lambda \otimes \mu', \lambda' \otimes \mu} \right) \enspace . 
\end{equation}
In the next section, we will show that if $\lambda, \mu, \lambda'$ and $\mu'$ have certain character-theoretic properties, then no irrep $\tau$ occurs too often in any of these natural distributions, and so the probability of a collision is small.

\section{Collisions, smoothness, and characters}
\label{sec:collisions}

Let us bound the probability $\Pcoll = \Pcoll^{\lambda \otimes \mu, \lambda' \otimes \mu'}$ that the natural distributions in $\lambda \otimes \mu$ and $\lambda' \otimes \mu'$ collide.  The idea is that $\Pcoll$ is small as long as both of either or both of these distributions is \emph{smooth}, in the sense that they are spread fairly uniformly across many $\tau$.  The following lemma shows that this notion of smoothness can be related to bounds on the normalized characters of typical representations.  

\begin{lemma}
\label{lem:coll}
Given a group $G$, say that an irrep $\lambda$ of $G$ is \emph{$f(n)$-smooth} if 
\[ \sum_{g \in G} \abs{\frac{\chi_\lambda(g)}{d_\lambda}}^4 \le f(n) \enspace . \]
Suppose that $\lambda$ and $\mu$ are $f(n)$-smooth.  Then
\[ \Pcoll \le \frac{\max_\tau d_\tau}{\sqrt{|G|}} \sqrt{f(n)} \enspace . \]
\end{lemma}

\begin{proof}
We have
\[ \Pcoll
= \sum_\tau P^{\lambda \otimes \mu}_\tau P^{\lambda' \otimes \mu'}_\tau 
\le \max_\tau \Plmt \enspace . \]
Using~\eqref{eq:ptau} and using Cauchy-Schwartz twice gives
\begin{align*}
\Plmt 
& = \frac{d_\tau}{d_\lambda d_\mu} \inner{\chi_\tau}{\chi_\lambda \chi_\mu}_G \\
& \le \frac{d_\tau}{d_\lambda d_\mu} 
\sqrt{ \innerg{\chi_\tau}{\chi_\tau} \innerg{\chi_\lambda \chi_\mu}{\chi_\lambda \chi_\mu} } \\
& = \frac{d_\tau}{d_\lambda d_\mu} 
\sqrt{ \innerg{\abs{\chi_\lambda}^2}{\abs{\chi_\mu}^2} } \\
& \le \frac{d_\tau}{d_\lambda d_\mu} 
\sqrt[4]{ \innerg{\abs{\chi_\lambda}^2}{\abs{\chi_\lambda}^2} \innerg{\abs{\chi_\mu}^2}{\abs{\chi_\mu}^2} } \\
& = \frac{d_\tau}{\sqrt{|G|}}
\sqrt[4]{ \sum_{g \in G} \abs{\frac{\chi_\lambda(g)}{d_\lambda}}^4 
\sum_{g \in G} \abs{\frac{\chi_\mu(g)}{d_\mu}}^4 } \\
& \le \frac{\max_\tau d_\tau}{\sqrt{|G|}} \sqrt{f(n)} \enspace .
\end{align*}
\end{proof}


Now let us focus on the case relevant to Graph Isomorphism, where $G=S_n$.  The reader can find an excellent introduction to the representation theory of $S_n$ in~\cite{JamesK1981}.  Here we recall that each irrep of the symmetric group $S_n$ corresponds to a \emph{Young diagram}, or equivalently an integer partition $\lambda_1 \ge \lambda_2 \ge \cdots$ where $\sum_i \lambda_i = n$.  The number of irreps is thus the partition number, 
\begin{equation} 
\label{eq:partition}
p(n) = (1+o(1)) \frac{1}{4\sqrt{3}\cdot n} \,e^{\delta \sqrt{n}} < e^{\delta \sqrt{n}} 
\end{equation}
where 
\[ \delta = \sqrt{2/3} \,\pi \enspace . \]
The maximum dimension of any irrep is bounded by the following result of Vershik and Kerov: 
\begin{theorem}[\cite{VershikK85}]
\label{thm:max-dimension}
  There is a constant $\hat{c} > 0$ such that for all $n \geq 1$,
  $$
  \max_\tau d_\tau 
  \leq e^{-(\hat{c}/2) \sqrt{n}} \sqrt{n!} \enspace .
  $$
\end{theorem}
\noindent
In this case, Lemma~\ref{lem:coll} gives 
\begin{equation}\label{eq:coll}
\Pcoll \le e^{-(\hat{c}/2) \sqrt{n}} \sqrt{f(n)} \enspace .
\end{equation}
Therefore, our goal is to show that typical irreps of $S_n$ are $f(n)$-smooth where $f(n)$ grows slowly enough with $n$.  Indeed, to prove our main result, even $f(n) = e^{a \sqrt{n}}$ for some $a < \hat{c}$ would suffice.  However, a well-motivated conjecture, described in the next section, implies $f(n)$-smoothness with $f(n) = O(\sqrt{n})$.

\section{A conjectured character bound}
\label{sec:conjecture}

In this section we present a conjecture regarding the high-dimensional irreps of $S_n$.  First we fix some notation.  Given a permutation $\pi \in S_n$, let $s(\pi)$ denote its \emph{support}, i.e., the number of objects not fixed by it.  Let $c(\pi)$ denote the number of nontrivial cycles of which $\pi$ is composed, i.e., other than fixed points, and let $t(\pi)$ denote the length of the shortest string of transpositions whose product is $\pi$.  Since a single $k$-cycle requires $k-1$ transpositions, we have $t(\pi) = s(\pi) - c(\pi)$.  

Then the following conjecture states that for all irreps whose dimension is sufficiently close to the maximum dimension, their normalized characters obey a certain uniform bound.

\begin{conjecture}
\label{conj}
Say that an irrep $\lambda$ of $S_n$ is \emph{big} if 
\[ d_\lambda > e^{-\sqrt{n} \log n} \sqrt{n!} \enspace . \]
Then there is a constant $A$ such that, for $n$ sufficiently large, the normalized character of all big $\lambda$ obeys
\[ \abs{\frac{\chi_\lambda(\pi)}{d_\lambda}} \le A^{t(\pi)} n^{-t(\pi)/2} \]
for all $\pi \in S_n$.
\end{conjecture}

We present evidence for this conjecture in Section~\ref{sec:evidence} below.  For now, we point out that it has the following two important consequences, which will imply our main result.

\begin{lemma}
\label{lem:bigsmooth}
Conjecture~\ref{conj} implies that all big irreps $\lambda$ are $O(\sqrt{n})$-smooth.
\end{lemma}

\begin{lemma}
\label{lem:inductivebig}
Say that an irrep $\lambda$ of $S_n$ is \emph{really big} if 
\[ d_\lambda > e^{-(1/2) \sqrt{n} \log n} \sqrt{n!} \enspace . \]
Suppose that $\lambda$ and $\mu$ are really big, and that $\tau$ is selected according to the natural distribution~\eqref{eq:ptau} in $\lambda \otimes \mu$.  Then Conjecture~\ref{conj} implies that $\tau$ is really big with probability $1-e^{-\omega(\sqrt{n})}$.
\end{lemma}

We will prove Lemmas~\ref{lem:bigsmooth} and~\ref{lem:inductivebig} with the help of the following combinatorial lemma.
\begin{lemma}
\label{lem:egf}
Let $0 < z < 1/n$, and let 
\[ q_n(z) = \sum_{\pi \in S_n} z^{t(\pi)} \enspace . \]
Then 
\begin{equation}
\label{eq:qn}
q_n(z) = O(\sqrt{n}) \,\frac{e^{-n}}{(1-zn)^{1/z}} \enspace .
\end{equation}
\end{lemma}

\begin{proof}
Consider the exponential generating function
\[ g(y,z) = \sum_{m=0}^\infty \frac{y^m}{m!} \,q_m(z) \enspace . \]
Using the techniques of~\cite[Chapter 3]{Wilf94}, we can write this as a product over all $k$ of contributions from the $(k-1)!$ possible $k$-cycles, including fixed points.  Since each such cycle contributes $k$ to $m$ and $k-1$ to $t(\pi)$, and since there are $(k-1)!$ $k$-cycles on a given set of $k$ objects, it follows that
\[ 
g(y,z) = \prod_{k=1}^\infty \exp\!\left( \frac{y^k z^{k-1}}{k} \right)
= \exp\!\left( \sum_{k=1}^\infty \frac{y^k z^{k-1}}{k} \right)
= \exp\!\left( -\frac{1}{z} \ln (1-yz) \right)
= \frac{1}{(1-yz)^{1/z}} \enspace . 
\] 
Now note that $e^{-n} g(n,z)$ is the expectation of $q_m(z)$, where $m$ is Poisson distributed with mean $n$.  Since $q_m(z) > 0$, this expectation is at least $q_n(z)$ times the probability that $m=n$, which is $e^{-n} n^n / n! = (1-o(1)) / \sqrt{2 \pi n}$.  Thus we have
\[ q_n(z) \le (1+o(1)) \sqrt{2 \pi n} \cdot e^{-n} g(n,z) \]
which concludes the proof.  
\end{proof}
Note that since the Poisson distribution is tightly peaked, if $q_n(z)$ varies smoothly with $n$ then $e^{-n} g(n,z)$ converges to $q_n(z)$ as $n \to \infty$.   In that case we could strengthen Lemma~\ref{lem:egf} by removing the $O(\sqrt{n})$ term; but, since this is not necessary to our result, we have not attempted to prove it.

\begin{proof}[Proof of Lemma~\ref{lem:bigsmooth}]
If the conjecture holds, then we have 
\begin{equation}
\label{eq:fqn}
\sum_{\pi \in S_n} \abs{\frac{\chi_\lambda(\pi)}{d_\lambda}}^4 
\le \sum_{\pi \in S_n} (A^{t(\pi)} n^{-t(\pi)/2})^4
= q_n(C/n^2) 
\end{equation}
where $C = A^4$.  Setting $z=C/n^2$ in Lemma~\ref{lem:egf} gives
\[ q_n(C/n^2) = O(\sqrt{n}) \,e^{-n} \left( 1-\frac{C}{n} \right)^{-n^2/C} \enspace . \]
Using $-\log (1-x) = x + x^2/2 + x^3/3 + \cdots$, we have 
\begin{align*}
e^{-n} \left( 1-\frac{C}{n} \right)^{-n^2/C} 
= \,e^{-n} \exp \left( \left(\frac{C}{n}+O\left( \frac{1}{n^2} \right) \right) 
  \left(\frac{n^2}{C} \right) \right)
= \,e^{-n} \,e^{n+O(1)}
= O(1) 
\end{align*}
which completes the proof.
\end{proof}

\begin{proof}[Proof of Lemma~\ref{lem:inductivebig}] 
Using simple dimension-counting arguments as in~\cite{Moore:2005:SGF}, the multiplicity $\innerg{\chi_\tau}{\chi_\lambda \chi_\mu}$ of $\tau$ in $\lambda \otimes \mu$ is at most $d_\tau$.  First we show that we can assume that $\tau$ is big; if it is not, by~\eqref{eq:ptau} we have
\begin{equation}
\label{eq:notbig}
 \Plmt \le \frac{d_\tau^2}{d_\lambda d_\mu} 
 < \frac{\left(e^{-\sqrt{n} \log n}\right)^2}{\left(e^{-(1/2) \sqrt{n} \log n} \right)^2}
 = e^{-\sqrt{n} \log n} = e^{-\omega(\sqrt{n})} \enspace . 
\end{equation}
Now, to bound the probability that $\tau$ fails to be \emph{really} big, we rewrite~\eqref{eq:ptau} as follows:
\[ \Plmt = \frac{d_\tau^2}{n!} 
\sum_{\pi \in S_n} \frac{\chi_\tau}{d_\tau} \frac{\chi_\lambda}{d_\lambda} \frac{\chi_\mu}{d_\mu} \]
(recall that the characters of $S_n$ are real).  Since $\tau$ is big, Conjecture~\ref{conj} applies to it as well as to $\lambda$ and $\mu$.  In that case we can bound $\Plmt$ using the triangle inequality, 
\[ \Plmt 
\le \frac{d_\tau^2}{n!} 
\sum_{\pi \in S_n} \abs{ \frac{\chi_\tau}{d_\tau} \frac{\chi_\lambda}{d_\lambda} \frac{\chi_\mu}{d_\mu} }
\le \frac{d_\tau^2}{n!} \sum_{\pi \in S_n} (A^{t(\pi)} n^{-t(\pi)/2})^3
= \frac{d_\tau^2}{n!} \,q_n(C/n^{3/2}) \]
where $C=A^3$.  Setting $z=C/n^{3/2}$ in Lemma~\ref{lem:egf} then gives
\[ q_n(C/n^{3/2}) = O(\sqrt{n}) \,e^{-n} \left( 1-\frac{C}{n^{1/2}} \right)^{-n^{3/2}/C} \enspace . \]
Again using $-\log (1-x) = x + x^2/2 + x^3/3 + \cdots$, we have 
\begin{align*}
e^{-n} \left( 1-\frac{C}{n^{1/2}} \right)^{-n^{3/2}/C}
&= \,e^{-n} \exp \left( \left(\frac{C}{n^{1/2}}+\frac{1}{2}\frac{C^2}{n}+O\left( \frac{1}{n^{3/2}} \right) \right) 
  \left(\frac{n^{3/2}}{C} \right) \right) \\
&= \,e^{-n} \,e^{n+(C/2)\sqrt{n}+O(1)} = O(e^{(C/2)\sqrt{n}}) \enspace . 
\end{align*}
The probability that $\tau$ fails to be really big, conditioned on the event that it is big, is thus less than 
\begin{equation}
\label{eq:notreallybig} 
\frac{d_\tau^2}{n!} \,O(\sqrt{n}) \,O(e^{(C/2)\sqrt{n}}) \,e^{\delta \sqrt{n}}
= e^{-\sqrt{n} \log n} \,e^{O(\sqrt{n})}
= e^{-\omega(\sqrt{n})} \enspace. 
\end{equation}
Combining~\eqref{eq:notbig} and~\eqref{eq:notreallybig}, the probability that $\tau$ fails to be really big is $e^{-\omega(\sqrt{n})}$.
\end{proof}

\section{Proof of the main result}
\label{sec:main}

We are now in a position to present our main result.

\begin{theorem}
\label{thm:main}
Let $\hat{c}$ be the constant of Theorem~\ref{thm:max-dimension}. Then Conjecture~\ref{conj} implies that for any constants $b,c$ such that $b < \hat{c}/2$ and $c < (\hat{c}/2) - b$, no sieve algorithm which combines less than $e^{b \sqrt{n}}$ coset states can solve Graph Isomorphism with success probability greater than $e^{-c \sqrt{n}}$.
\end{theorem}

\begin{proof}
Let us first consider the behavior of a sieve algorithm $A$ in the case where the hidden subgroup $H \subset S_n \wr \Z_2$ is trivial.  For convenience, extend the notion of ``really big'' to representations of $S_n \wr \Z_2$ by declaring $\sigma_{\{\lambda, \mu\}}$ to be really big when both $\lambda$ and $\mu$ are.  We will establish that with overwhelming probability, all the irrep labels observed by $A$ are both really big and inhomogeneous.  

Let $\ell < e^{b \sqrt{n}}$ be the number of coset states initially generated by the algorithm.  We begin by showing that with high probability, the irrep labels on the $\ell$ leaves, i.e., those resulting from weak Fourier sampling, are all really big and homogeneous.  If $H$ is trivial, then these irrep labels are Plancherel-distributed, and by~\eqref{eq:planch-wreath-inhom} the probability that a given one fails to be really big is at most twice the probability that a Plancherel-distributed irrep of $S_n$ fails to be.  The Plancherel measure of the set of non-really-big irreps of $S_n$ is at most the number of irreps times the measure of a single such $\lambda$, so by~\eqref{eq:partition} this probability is at most
\[ 2 p(n) \frac{d_\lambda^2}{n!} 
< 2 e^{\delta \sqrt{n}} e^{-\sqrt{n} \log n} 
= e^{-\omega(\sqrt{n})} \enspace . \]
Moreover, by~\eqref{eq:planch-wreath-hom} the probability that the label of a given leaf is homogeneous is the probability that we observe the same irrep of $S_n$ in two samples of the Plancherel distribution.  Using Theorem~\ref{thm:max-dimension} this is
\[ \sum_{\lambda} \left( \frac{d_\lambda^2}{|n!|} \right)^{\!2}
< \max_\lambda \frac{d_\lambda^2}{|n!|}
\le e^{-\hat{c} \sqrt{n}} \enspace . \]
Thus the combined probability that any of the $\ell$ leaves have a label which is not both really big and inhomogeneous is
\begin{equation}
\label{eq:event1}
\ell \left( e^{-\omega(\sqrt{n})} + e^{-\hat{c} \sqrt{n}} \right) < e^{(b-\hat{c}+o(1)) \sqrt{n}} \enspace .
\end{equation}

Now, assume inductively that all the irreps observed by the algorithm before the $i$th combine-and-measure step are really big and inhomogeneous, and that the $i$th step combines states with two such labels $\sigma_{\{\lambda,\lambda'\}}$ and $\sigma_{\{\mu,\mu'\}}$.  Recall from~\eqref{eq:prob-hom-bound} that the probability this results in a homogeneous irrep is bounded by the probability $\Pcoll$ of a collision between a pair of natural distributions in $S_n$.  Then Conjecture~\ref{conj}, Lemmas~\ref{lem:coll} and~\ref{lem:bigsmooth}, and Theorem~\ref{thm:max-dimension} imply that this probability is bounded by
\[ \Pcoll \le e^{-(\hat{c}/2) \sqrt{n}} \,O(n^{1/4}) \enspace . \]
In addition, Conjecture~\ref{conj} and Lemma~\ref{lem:inductivebig} imply that the the probability the observed irrep fails to be really big is $e^{-\omega(\sqrt{n})}$.  Since each combine-and-measure step reduces the number of states by one, there are less than $\ell$ such steps; taking a union bound over all of them, the probability that any of the observed irreps fail to be both homogeneous and really big is 
\begin{equation}
\label{eq:event2}
 \ell \left( e^{-(\hat{c}/2) \sqrt{n}} \,O(n^{1/4}) + e^{-\omega(\sqrt{n})} \right) 
< e^{(b-(\hat{c}/2)+o(1)) \sqrt{n}} \enspace . 
\end{equation}

Let us call a transcript inhomogeneous if all of its irrep labels are.  As $c < (\hat{c}/2) - b$ by assumption, combining~\eqref{eq:event1} and~\eqref{eq:event2}, for $n$ sufficiently large, $A$'s transcript is inhomogeneous with probability at least $1-e^{-c \sqrt{n}}$.

Now consider $A$'s behavior in the case of a nontrivial hidden subgroup, $H = \{1,m\}$.  Inductively applying Equation~\eqref{eq:equality} shows that the probability of observing any inhomogeneous transcript is exactly the same as it would have been if $H$ were trivial. Thus the total variation distance between the distribution of transcripts generated by $A$ in these two cases is no more than $e^{-c \sqrt{n}}$, and the theorem is proved.
\end{proof}

\section{Evidence for the conjecture}
\label{sec:evidence}

In this section we give some evidence for our conjecture, by showing that it holds in a variety of cases.  First we prove the following lemma, which bounds the height and width of a Young diagram corresponding to a big representation.

\begin{lemma}
\label{lem:width}
If $\lambda$ is big, then the width and height of its Young diagram are less than $4 \sqrt{n} \log n$.
\end{lemma}

\begin{proof}
The Robinson-Schensted correspondence~\cite{Fulton97} maps permutations to Young diagrams in such a way that the uniform measure on $S_n$ maps to the Plancherel measure.  In addition, the width (resp.\ height) of the Young diagram is equal to the length of the longest increasing (resp.\ decreasing) subsequence.  

It follows that if $\lambda$ has width $w$, its Plancherel measure $d_\lambda^2/n!$ is at most the probability that a random permutation in $S_n$ has an increasing subsequence of length $w$.  By Markov's inequality, this probability is at most the expected number of such subsequences, which is 
\[ {n \choose w} \frac{1}{w!} < \frac{n^w}{w!^2} < \left( \frac{e \sqrt{n}}{w} \right)^{\!2w} \]
since $w! > w^w e^{-w}$.  If $w = 4 \sqrt{n} \log n$, comparing this to the Plancherel measure of a big irrep $\lambda$ gives 
\[ 
e^{-2 \sqrt{n} \log n} 
< \frac{d_\lambda^2}{n!} 
\le \left( \frac{e}{4 \log n} \right)^{8 \sqrt{n} \log n}
< (4/e)^{-8 \sqrt{n} \log n} 
\]
and since $e < (4/e)^4$ this is a contradiction.  The argument for the height is identical.
\end{proof}

Our next lemma uses deep results of Kerov~\cite{Kerov03} and Biane~\cite{Biane98} to show that our conjecture holds for permutations whose support grows slowly enough with $n$.  Indeed, it was Biane's work, which builds on recent results in the theory of noncommuting random variables, that originally inspired our conjecture.

\begin{lemma}
\label{lem:constant-support}
Conjecture~\ref{conj} holds for all permutations whose support is less than $(\log n)^{1-\beta}$ for some $\beta > 0$.
\end{lemma}

\begin{proof}  We first use a beautiful fact due to Kerov~\cite{Kerov03}.  Any sequence of irreps $\lambda_n$ of $S_n$ whose dimensions differ less than exponentially from the maximum dimension as $n \to \infty$, that is, such that $d_{\lambda_n} = e^{-o(n)} \sqrt{n!}$, have Young diagrams which converge to an asymptotic shape when scaled by $\sqrt{n}$.  This condition is clearly satisfied by any sequence of big irreps; for aesthetic reasons, we show this asymptotic shape in Figure~\ref{fig:kerov}. 

Note that these Young diagrams have width and height $2\sqrt{n}$, thus solving Ulam's celebrated problem of the likely length of the longest increasing or decreasing subsequence in a random permutation.  However, the notion of convergence used in~\cite{Kerov03} does not rule out the possibility that the Young diagrams corresponding to big $\lambda$ have long thin ``tails'' near the horizontal and vertical axes, so we rely on Lemma~\ref{lem:width} to bound the length of these tails. 

\begin{figure}[htb]
  \begin{center}
    \includegraphics[width=0.35\columnwidth]{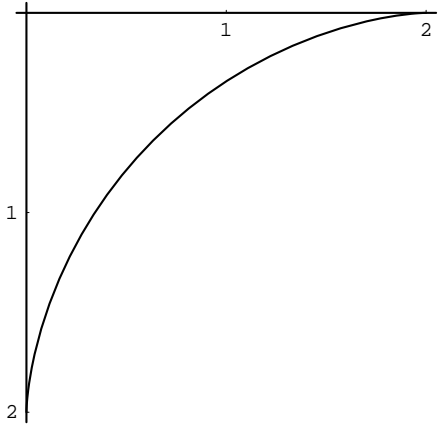}
  \end{center}
\caption{The asymptotic shape of a Young diagram associated with an irrep $\lambda$ of $S_n$ whose dimension is close to the maximum dimension.  It is scaled by $\sqrt{n}$, so its height and width are $2\sqrt{n}$.}
\label{fig:kerov}
\end{figure}

Now, Theorem 1.3 of~\cite{Biane98} and the surrounding discussion establishes that for any sequence of irreps $\lambda$ whose Young diagrams converge to some asymptotic shape, and which moreover have height and width $o(n^{1/2 + \eps})$ for all $\eps > 0$ (which follows in our case from Lemma~\ref{lem:width}) the normalized character of any permutation with small support can be approximated by a product over its cycles.  Specifically, for any asymptotic shape there is a set of constants $a_k$ for $k \ge 1$, called the \emph{free cumulants} associated with that shape,\footnote{Readers should be warned that our notation differs somewhat from Biane's.} such that, if $s(\pi) < (\log n)^{1-\beta}$ for some fixed $\beta > 0$, 
\[ \frac{\chi_\lambda(\pi)}{d_\lambda} = n^{-t(\pi)/2} \prod_{k \ge 1} a_k^{c_k(\pi)} + o(n^{-t(\pi)/2}) \] 
where $c_k(\pi)$ is the number of cycles of size $k$ in $\pi$.  For the Kerov shape in particular, $a_1 = 1$ and $a_k = 0$ for all $k \ge 2$, giving a bound even tighter than our conjecture, 
\[ \frac{\chi_\lambda(\pi)}{d_\lambda} = o(n^{-t(\pi)/2}) \] 
for any $\pi$ other than the identity.
\end{proof}
\noindent

The next two lemmas use simple counting arguments to show that Conjecture~\ref{conj} holds for permutations of large enough support, or which contain large cycles.

\begin{lemma}
\label{lem:large-support}
Conjecture~\ref{conj} holds for all $\pi$ with $t(\pi) > \sqrt{n} \log n$.
\end{lemma}

\begin{proof}
We use the Murnaghan-Nakayama formula for the character~\cite{JamesK1981}.  A \emph{ribbon tile} of length $k$ is a polyomino of $k$ cells, arranged in a path where each step is up or to the right.  Given a Young diagram $\lambda$ and a permutation $\pi$ with cycle structure $k_1 \ge k_2 \ge \cdots$, a \emph{consistent tiling} consists of removing a ribbon tile of length $k_1$ from the boundary of $\lambda$, then one of length $k_2$, and so on, with the requirement that the remaining part of $\lambda$ is a Young diagram at each step.  Let $h_i$ denote the height of the ribbon tile corresponding to the $i$th cycle: then the Murnaghan-Nakayama formula states that
\begin{equation}
\label{eq:m-n}
 \chi_\lambda(\pi) = \sum_T \prod_i (-1)^{h_i+1} 
\end{equation}
where the sum is over all consistent tilings $T$.  

\begin{figure}[htb]
  \begin{center}
    \includegraphics[width=0.4\columnwidth]{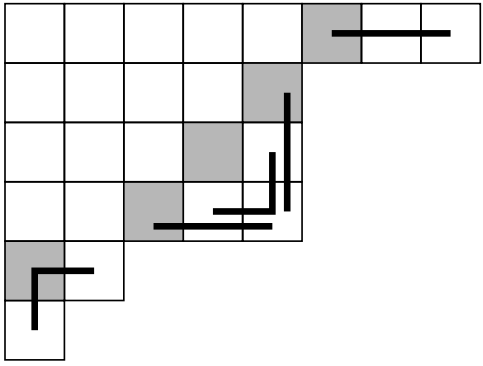}
  \end{center}
\caption{We associate each possible location for a ribbon tile of fixed length $k$ with a cell (shaded) which is above the tile's lower end and to the left of its upper end.  The resulting sequence of cells moves up and to the right at each step, implying that the number of locations is less than $\sqrt{2n}$.  Here $k=3$.}
\label{fig:locations}
\end{figure}

Clearly the number of consistent tilings is an upper bound on $\abs{\chi_\lambda(\pi)}$.  Now, we claim that for any fixed $k$, the number of possible locations for a ribbon tile of length $k$ on the boundary of a Young diagram $\lambda$ of size $n$ is less than $\sqrt{2 n}$.  To see this, associate each one with the cell of $\lambda$ which is directly above the tile's lower end, and directly to the left of its upper end, as shown in Figure~\ref{fig:locations}.  A little thought reveals that the resulting sequence of cells has the property that each one is above and to the right of the previous one.  Therefore, if there are $\ell$ locations, we have
\[ n \ge \sum_{i=1}^\ell i > \ell^2 / 2 \enspace . \]
and so $\ell < \sqrt{2n}$.  It follows that the number of ways to remove the ribbon tiles corresponding to the $c(\pi)$ nontrivial cycles is less than  
\[ (2n)^{c(\pi)/2} \enspace . \]
Moreover, after these ribbon tiles are removed, the number of consistent tilings of the remaining Young diagram is simply the dimension of the corresponding irrep of $S_{n-s(\pi)}$, which is less than  $\sqrt{|S_{n-s(\pi)}|} = \sqrt{(n-s(\pi))!}$.  Therefore, if $\lambda$ is big we have
\begin{align*} 
\abs{\frac{\chi_\lambda(\pi)}{d_\lambda}} 
&< \frac{(2n)^{c(\pi)/2} \sqrt{(n-s(\pi))!}}{e^{-\sqrt{n} \log n} \sqrt{n!}} \\
&< 2 \cdot e^{\sqrt{n} \log n} \,2^{c(\pi)/2} \,e^{s(\pi)/2} \,n^{(c(\pi)-s(\pi))/2} \\
&\le 2 \cdot e^{\sqrt{n} \log n} \,(\sqrt{2} e)^{t(\pi)} \,n^{-t(\pi)/2} \enspace .
\end{align*}
Here we used the bound $(n-s)!/n! < 4 \cdot n^{-s} e^s$, implied by Stirling's approximation,  
and the facts that $c(\pi) \le t(\pi)$, $s(\pi) \le 2 t(\pi)$, and $t(\pi) = s(\pi) - c(\pi)$.  Finally, if $t(\pi) > \sqrt{n} \log n$, the term $e^{\sqrt{n} \log n}$ can be absorbed into $A^{t(\pi)}$, and Conjecture~\ref{conj} holds for any $A > \sqrt{2} e^2$.
\end{proof}

The next lemma shows that Conjecture~\ref{conj} also holds for permutations that contain sufficiently large cycles, for the simple reason that in this case the character is zero.

\begin{lemma}
\label{lem:big-cycles}
Conjecture~\ref{conj} holds for all $\pi$ that contain a cycle of size $8 \sqrt{n} \log n$ or greater.
\end{lemma}

\begin{proof}  The maximum size of any ribbon tile that can be placed on the boundary of a Young diagram is its width plus its height (minus one).  Thus if such a cycle exists, by Lemma~\ref{lem:width}   there are no consistent tilings and the Murnaghan-Nakayama formula~\eqref{eq:m-n} implies that the character is zero.
\end{proof}

Unfortunately, Lemmas~\ref{lem:constant-support}, \ref{lem:large-support}, and \ref{lem:big-cycles} leave uncovered the case of permutations whose support is between, roughly, $\log n$ and $\sqrt{n} \log n$ and which consist entirely of short cycles.  However, it seems very plausible to us that the scaling we conjecture here extends across this gap.  

Moreover, it should be noted our main result still holds even if the $\log n$ term in our definition of big representations were replaced by a more slowly-growing function of $n$ such as $\log \log n$.  Thus we can get by with significantly weaker forms of the conjecture, which apply only to representations even closer to the maximum dimension than we consider here.

\section*{Acknowledgments}

We are very grateful to Philippe Biane, Persi Diaconis, and Piotr \'Sniady for helpful conversations about the character theory of the symmetric groups, and to Sally Milius, Tracy Conrad and Rosemary Moore for inspiration.  This work was supported by NSF grants CCR-0220070, EIA-0218563, and CCF-0524613, and ARO contract W911NF-04-R-0009.  


\newcommand{\etalchar}[1]{$^{#1}$}
\ifx \k \undefined \let \k = \c \immediate\write16{Ogonek accent unavailable:
  replaced by cedilla} \fi\ifx \ocirc \undefined \def \ocirc
  #1{{\accent'27#1}}\fi\ifx \mathbb \undefined \def \mathbb #1{{\bf #1}}
  \fi\ifx \mathbb \undefined \def \mathbb #1{{\bf #1}}\fi

\end{document}